# SOJOURN TIMES FOR THE ONE DIMENSIONAL GROVER WALK

**CLEMENT AMPADU**

31 Carrolton Road
Boston, Massachusetts, 02132
USA
e-mail: drampadu@hotmail.com

## Abstract

Using the technique of path counting we show non-existence of sojourn times in the Grover walk which is related to the Grover's algorithm in computer science.

## I. Introduction

The study of sojourn times has been a topic of intensive study in the literature. Concerning sojourn times for stochastic processes, in [1] the authors investigate the sojourn time above a high threshold of a continuous stochastic process on the unit interval. In [2] the authors study a high-order heat –type equation associated with a Markov-pseudo process, and study the sojourn time in the interval $[0, \infty)$ up to a fixed time for the process. In [3] the authors study the asymptotics of the stationary sojourn time on the integers of a "typical customer" in a tandem of single-server queues. It is shown that, in a certain "intermediate" region of light tailed service time distributions on the integers, the sojourn time on the integers may take a large value mostly due to a large value of a single service time of one of the customers. In [4] the authors study the M/M/1 queue with processor sharing. The conditional sojourn time distribution is studied, conditioned on the customers service requirement, in various limits. A processor shared M/M/1 queue that can accommodate at most a finite number K of customers is studied in [5]. An exact expression for the sojourn time distribution in the finite capacity model, in terms of a Laplace transform is given. In [6] the M/M/1-PS queue with processor sharing is considered. The conditional sojourn time distribution of an arriving customer, conditioned on the number of other customers present is studied, as well as their asymptotic limits. A new formula is obtained for the conditional sojourn time distribution, using a discrete Green's function. In [7] the M/G/1 queue with processor sharing server is considered. The conditional sojourn time distribution is studied, conditioned on the customer's service requirement, as well as the unconditional distribution, in various asymptotic limits. In [8] processor sharing queues are studied, the author obtains new formulas for the moments of the stationary sojourn time in the M/G/1-EPS queue with a finite number of jobs of infinite size. The asymptotic decay rate of the sojourn time of a customer in the stationary M/G/1 queue under the Foreground-Background (FB) service discipline is studied in [9]. The authors prove that for light-tailed service times the decay rate of the sojourn time is equal to the decay rate of the busy period. It is shown that FB minimizes the decay rate in the class of working-conserving disciplines. In [10], a general method of calculating conditional sojourn times based on stochastic absorption is given, the case of a delta dimer is considered, in showing how to correct quantum clocks using optical potential methods. The question of whether the quantum-mechanical sojourn time can be clocked without the clock affecting the sojourn time is studied in [11]. It is shown that the non-unitary clock involving the imaginary potential can lead to a negative conditional sojourn time for non-random potentials.

As for sojourn times for random walks, in [12] the continuous time random walk described by arbitrary sojourn time probability density functions is studied leading to general expressions for the distribution of time-averaged observables for such systems. In [13] the probability distributions of the sojourn time related to the Bernoulli random walk on the integers is studied. By considering the discrete counterpart to the famous Paul Levy arcsine law for Brownian motion, the authors give a representation for the probability distribution related to the random walk subject to possible conditioning. The main tool is the use of generating functions which is also considered in the present paper.

The sojourn times for quantum walks has not been a subject of intensive study in the literature, to our knowledge the *founding paper* appears to be due to Konno [14] and the present paper is the *second* to study sojourn times for quantum walks. In [14] the sojourn time for the quantum walk on the line governed by the Hadamard operator is studied by the method of path counting leading to generating functions for the sojourn times. In this paper we consider the Grover operator instead.

The rest of this paper is organized as follows. Section II gives the definition of our model.

In Section III we present the main results with proof via Theorem 7 and Theorem 8, which show that the complex power series representation of the sojourn times are divergent. Section IV is devoted to an open problem concerning classifying matrices up to general dimensionality with the property *non-existence of sojourn times imply localization*.

**II.   Definition of the Walk**

Recall that the discrete-time quantum walk is the quantum analogue of the classical random walk with an additional degree of freedom called the chirality. The chirality takes values left and right, and it means the direction of the motion of the walker. At each time step, if the walker has left chirality, it moves one step to the left, and if it has right chirality it moves one step to the right.

In this paper we put $|L\rangle = \begin{bmatrix} 1 \\ 0 \end{bmatrix}$ and $|R\rangle = \begin{bmatrix} 0 \\ 1 \end{bmatrix}$ where $L$ and $R$ refer to the left and right chirality state respectively. Suppose that the time evolution of the walk is governed by the following general unitary matrix $U = \begin{bmatrix} a & b \\ c & d \end{bmatrix}$, where the entries are complex numbers. To describe the dynamics of the walk, write $U = P + Q$, where $P = \begin{bmatrix} a & b \\ 0 & 0 \end{bmatrix}$ and $Q = \begin{bmatrix} 0 & 0 \\ c & d \end{bmatrix}$. Note that the matrix $P$ represents the left moving walker, whilst the matrix $Q$ represents the right moving walker,

at the position $x$ on the line at each time step, respectively.

The Grover operator as is well known was first introduced by Moore and Russell in their study of quantum walks on the hypercube [15]. Based on Grover's diffusion, the operator has elements $a_{ij} = \frac{2}{d} - \delta_{ij}$, where $\delta_{ij} = \begin{cases} 1, i = j \\ 0, i \neq j \end{cases}$ and $d = 2^{\tilde{d}}$ where $\tilde{d}$ is the lattice dimension of the quantum walk. The Grover walk is related to the Grover algorithm in the computer science. In the present paper we focus on $U = G$:

$$G = \begin{bmatrix} 0 & 1 \\ 1 & 0 \end{bmatrix}.$$ We will take the initial quibit state as $\varphi^* = \begin{bmatrix} 0 & i \end{bmatrix}^T$, where $T$ is the transpose operator. Let $\Xi_n(l,m)$ denote the sum of all paths starting from the origin in the trajectory consisting of $l$ steps left, and $m$ steps right. In fact, for time $n = l + m$, and position $x = -l + m$ we have $\Xi_n(l,m) = \sum_{l_j, m_j} P^{l_1} Q^{m_1} P^{l_2} Q^{m_2} \ldots P^{l_{n-1}} Q^{m_{n-1}} P^{l_n} Q^{m_n}$, where the summation is taken over all integers $l_j, m_j \geq 0$ satisfying $\sum_{j=1}^{n} l_j = l$, $\sum_{j=1}^{n} m_j = m$, and $l_j + m_j = 1$. We should note that the definition gives $\Xi_{n+1}(l,m) = P\Xi_n(l-1,m) + Q\Xi_n(l,m-1)$. The probability that the our quantum walker is in position $x$ at time $n$ starting from the origin with $\varphi^* \left( = \begin{bmatrix} 0 & i \end{bmatrix}^T \right)$ is defined by

$$P(X_n = x) = \left\| \Xi_n(l,m)\varphi^* \right\|^2$$ where $n = l + m$ and $x = -l + m$.

Let $\Psi_n^x(k)$ be the sum of all paths that the quantum walker starting from $x$ spends exactly $k$ intervals of time to the right of the origin, where $\Psi_n^x(k) = \sum_{y=-n+x}^{n+x} \Psi_n^{x \to y}(k)$, and $\Psi_n^{x \to y}(k)$ denotes the sum of all paths that the quantum walker, which starts at position $x$ and reaches position $y$ at time $n$, spends exactly $k$ intervals of time to the right of the origin. Also let $\Gamma_n(k)$ be the sum of all paths that the quantum walker starting from the origin and returning to the point spends exactly $k$ intervals of time, up to time $n$, to the right of the origin. We give the generating functions of

$\Psi_n^x(k)$ and $\Gamma_n(k)$ in the next section.

### III. Main Results

To give the generating function of $\Psi_n^x(k)$ for the Grover walk, we introduce the following matrices

$$R = \begin{bmatrix} 0 & 1 \\ 0 & 0 \end{bmatrix}, \ S = \begin{bmatrix} 0 & 0 \\ 0 & 1 \end{bmatrix}, \ P = \begin{bmatrix} 0 & 1 \\ 0 & 0 \end{bmatrix}, \ Q = \begin{bmatrix} 0 & 0 \\ 1 & 0 \end{bmatrix}.$$ Notice that $P, Q, R, S$ form an orthonormal basis of the vector space of complex $2 \times 2$ matrices with respect to the trace inner product

$\langle A | B \rangle = tr(A^*B)$, where * means the adjoint operator. So we can write $\Psi_n^x(k)$ as a linear combination of $P, Q, R, S$ say $\Psi_n^x(k) = p_n^x(k)P + q_n^x(k)Q + r_n^x(k)R + s_n^x(k)S$. For $u \in \{p, q, r, s\}$ we introduce the generating function $\tilde{u}^x(z,t) = \sum_{k=0}^{\infty} \sum_{n=1}^{\infty} u_n^x(k) z^n t^k$, and write

$\bar{u}^x(z,t) = \tilde{u}^x(z,t) + \tilde{u}^x(-z,t) + \tilde{u}^x(z,-t) + \tilde{u}^x(-z,-t)$, we see that $\bar{u}^x(z,t)$ makes sense and is non-vanishing if and only if both $k, n$ are even so we can write $\bar{u}^x(z,t) = \sum_{k=0}^{\infty} \sum_{n=1}^{\infty} u_n^x(2k) z^{2n} t^{2k}$.

For the Grover walk, by combining Lemma 4.2 and Lemma 4.4 in Konno [14], we have the following

**Lemma 1 (Towards $\bar{p}^0(z,t)$ and $\bar{r}^0(z,t)$):**

$$\frac{zt\lambda_1}{1 - zt\lambda_1} C_1^{(p)} = C_1^{(r)}, \ \frac{\lambda_2}{z} C_2^{(p)} = C_2^{(r)}$$

$$C_1^{(p)} - \frac{zt}{\phi(zt)} = C_2^{(p)} - \frac{z}{\phi(z)} \left( = \tilde{p}^0(z,t) \right), \ C_1^{(r)} - \frac{(zt)^2}{\phi(zt)} = C_2^{(r)} - \frac{z^2}{\phi(z)} \left( = \tilde{r}^0(z,t) \right)$$

**Lemma 2 (Towards $\bar{q}^0(z,t)$ and $\bar{s}^0(z,t)$):**

$$C_2^{(q)} - \frac{z}{\phi(z)} = zt\left[C_1^{(s)}\lambda_1 - \frac{(zt)^2}{\phi(zt)} + 1\right], \quad C_2^{(s)} - \frac{z^2}{\phi(z)} = z\left[\frac{C_2^{(q)}}{\lambda_2} - \frac{z}{\phi(z)}\right]$$

$$C_2^{(q)} - \frac{z}{\phi(zt)} = C_1^{(q)}\left(= \tilde{q}^0(z,t)\right), \quad C_1^{(s)} - \frac{(zt)^2}{\phi(zt)} = C_2^{(s)} - \frac{z^2}{\phi(z)}\left(= \tilde{s}^0(z,t)\right)$$

From Lemma 1 we can deduce the following.

**Lemma 3 ( Towards $\bar{p}^0(z,t)$ and $\bar{r}^0(z,t)$):**

$$\tilde{r}^0(z,t) = \frac{zt\lambda_1}{1-zt\lambda_1}C_1^{(p)}, \quad \tilde{p}^0(z,t) = C_1^{(p)} - \frac{zt}{\phi(zt)}, \text{ where}$$

$$C_1^{(p)} = \left[\frac{zt}{1-zt\lambda_1} - \frac{\lambda_2}{z}\right]^{-1} \times \left\{\left[\frac{(zt)^2}{\phi(zt)} - \frac{z^2}{\phi(z)}\right] - \frac{\lambda_2}{z}\left[\frac{zt}{\phi(zt)} - \frac{z}{\phi(z)}\right]\right\}$$

From Lemma 2 we can deduce the following.

**Lemma 4 (Towards $\bar{q}^0(z,t)$ and $\bar{s}^0(z,t)$):**

$$\tilde{s}_0(z,t) = \frac{zC_2^{(q)}}{\lambda_2} - \frac{z^2}{\phi(z)}, \quad \tilde{q}_0(z,t) = C_2^{(q)} - \frac{z}{\phi(z)}, \text{ where}$$

$$C_2^{(q)} = \left[1 - \frac{z^2 t\lambda_1}{\lambda_2}\right]^{-1} \times \left\{\frac{z - z^3 t\lambda_1}{\phi(z)} + \frac{(zt)^3 t\lambda_1 - (zt)^3}{\phi(zt)} + zt\right\}$$

**Lemma 5:** $\bar{p}^0(z,t) = \tilde{p}^0(z,t) + \tilde{p}^0(-z,t) + \tilde{p}^0(z,-t) + \tilde{p}^0(-z,-t)$,

$\bar{q}^0(z,t) = \tilde{q}^0(z,t) + \tilde{q}^0(-z,t) + \tilde{q}^0(z,-t) + \tilde{q}^0(-z,-t)$,

$\bar{s}^0(z,t) = \tilde{s}^0(z,t) + \tilde{s}^0(-z,t) + \tilde{s}^0(z,-t) + \tilde{s}^0(-z,-t)$,

$\bar{r}^0(z,t) = \tilde{r}^0(z,t) + \tilde{r}^0(-z,t) + \tilde{r}^0(z,-t) + \tilde{r}^0(-z,-t)$, where $\tilde{r}^0(z,t)$ and $\tilde{p}^0(z,t)$ are given by Lemma 3; $\tilde{q}^0(z,t)$ and $\tilde{s}^0(z,t)$ are given by Lemma 4.

Now we give the generating function of $\Gamma_n(k)$. We start with following crucial statement whose proof is similar in nature to Lemma 5.1 in Konno[14], therefore we omit it.

**Lemma 6:** For $k = 0, 1, \cdots, n$ we have

$$\Gamma_{2n}(2k) = I_{\{1,2,\cdots,k\}}(k) \times \sum_{r=1}^{k} \Gamma_{2n-2r}(2k-2r)\Gamma_{2r}(2r) + I_{\{0,1,\cdots,n-1\}}(k) \times \sum_{r=1}^{n-k} \Gamma_{2n-2r}(2k)\Gamma_{2r}(0)$$

where $I_A(k)$ is the indicator function of a set $A$ and

$$\Gamma_{2r}(2r) = \frac{a_{2r-1}}{2}\begin{bmatrix} 0 & 1 \\ 0 & 0 \end{bmatrix} \text{ and } \Gamma_{2r}(2r) = \frac{a_{2r-1}}{2}\begin{bmatrix} 0 & 0 \\ -1 & 0 \end{bmatrix} \quad (r \geq 1), \text{ with } \sum_{n=1}^{\infty} a_n z^n \text{ divergent}$$

If we put $\overline{\Gamma}(z,t) = \sum_{n=1}^{\infty}\sum_{k=1}^{\infty} \Gamma_{2n}(2k) t^{2k} z^{2n}$, then by Lemma 6 we have $\overline{\Gamma}_{2n}(t) = (\overline{\Gamma}_{2n}(t) + I)X$, where $I$ is the $2 \times 2$ identity matrix, and $X = \sum_{r=1}^{\infty}\{\Gamma_{2r}(2r)(zt)^{2r} + \Gamma_{2r}(0)z^{2r}\}$. Now by $\overline{\Gamma}_{2n}(t) = (\overline{\Gamma}_{2n}(t) + I)X$, we have $\overline{\Gamma}_{2n}(z,t) = X(I-X)^{-1}$, since $\sum_{n=1}^{\infty} a_n z^n$ is divergent, it follows that $X$ is the matrix with infinite components. In particular we have the following.

**Theorem 7:** $\overline{\Gamma}(z,t) = divergent$

We should remark that $\sum_{n=1}^{\infty} a_n z^n$ is divergent stems from the supposition that the unitary matrix in this paper is the one-dimensional Grover walk, where the first entry in the matrix is zero, it implies that $\lambda_{\pm}(z) = \infty$ in Konno [14]. In particular we also see that we have the following from Lemma 5

**Theorem 8:** $\overline{p}^0(z,t) = \overline{q}^0(z,t) = \overline{s}^0(z,t) = \overline{r}^0(z,t) = divergent$.

### IV. Open Problem

For the Grover walk considered in this paper, the unitary transformation in the Fourier space is given by $G(k) = \begin{pmatrix} 0 & e^{-ik} \\ e^{ik} & 0 \end{pmatrix}$, whose eigenvalues are both independent of $k$. Recall that that the degeneracy of the eigenvalues is a necessary condition for localization, hence localization occurs in the Grover transformation considered in this paper. Since it is a well known fact that the generating function completely determines the probability distribution, it is an interesting problem to investigate the relationship between sojourn times and the existence of localization. Based on our work we make the following.

**Conjecture :** Non-existence of sojourn times imply localization.

Concerning the conjecture, for quantum walks on the line governed by unitary matrices of the form $\left\{ \begin{bmatrix} 0 & b \\ c & 0 \end{bmatrix} : b, c \in C \right\}$, non-existence of Sojourn times is certain, also in the Fourier space the unitary transformation becomes $\begin{bmatrix} 0 & be^{-ik} \\ ce^{ik} & 0 \end{bmatrix}$ whose eigenvalues are independent of $k$, hence localization occurs.

So the question is whether the conjecture holds for unitary matrices of general dimensionality. Can we classify all such matrices up to dimensionality?